\documentclass{sig-alternate}
\permission{Copyright is held by the International World Wide Web Conference Committee (IW3C2). IW3C2 reserves the right to provide a hyperlink to the author's site if the Material is used in electronic media.}
\conferenceinfo{WWW'14,}{April 7--11, 2014, Seoul, Korea.} 
\copyrightetc{ACM \the\acmcopyr}
\crdata{978-1-4503-2744-2/14/04}

\usepackage{url}
\usepackage{multirow}

\newcommand{\argmax}{\operatornamewithlimits{argmax}}

\usepackage{times}
\usepackage{latexsym}
\usepackage{amsmath}
\usepackage{multirow}
\usepackage{graphicx}
\usepackage{multirow}
\usepackage{subfigure}
\usepackage{url}

\usepackage{color}
\usepackage{url}
\usepackage{enumitem}

\usepackage{flushend}

\newenvironment{tightitemize}%
  {\begin{itemize}[topsep=0pt, partopsep=0pt] \footnotesize%
    \setlength{\itemsep}{0pt}%
    \setlength{\parskip}{0pt}%
    }%
  {\end{itemize}}

  {\begin{itemize}[topsep=10pt, partopsep=0pt] \footnotesize%
    \setlength{\itemsep}{2pt}%
    \setlength{\parskip}{0pt}%
    }%
  {\end{itemize}}

\begin{document}

\title{Timeline Generation: Tracking individuals on Twitter}

\numberofauthors{2} 
\author{
\alignauthor
Jiwei Li\\
       \affaddr{School of Computer Science}\\
      \affaddr{Carnegie Mellon University}\\
       \affaddr{Pittsburgh, PA 15213}\\
       \affaddr{bdlijiwei@gmail.com}
\alignauthor
Claire Cardie\\
       \affaddr{Department of Computer Science}\\
       \affaddr{Cornell University}\\
       \affaddr{Ithaca, NY 14850}\\
       \affaddr{cardie@cs.cornell.edu}
}
\maketitle

\begin{abstract}
In this paper, we preliminarily learn the problem of
reconstructing users' life history based on the their Twitter stream
and proposed an unsupervised framework that
create a chronological list for {\it personal important events} (PIE)
of individuals.
By analyzing individual tweet collections, we find that what are suitable for inclusion in the personal timeline should be tweets talking about
personal (as opposed to public) and time-specific (as opposed to time-general) topics. 
To further extract these types of topics,
we introduce a non-parametric multi-level Dirichlet Process model
to recognize four
types of tweets: personal time-specific (PersonTS), personal time-general
(PersonTG), public time-specific (PublicTS) and public time-general
(PublicTG) topics, which, in turn, are used for further personal event
extraction and timeline generation. 
To the best of our knowledge, this is the first work focused on the generation of timeline for individuals from Twitter data. 
For evaluation, we have built gold standard timelines
that contain PIE related events from 20 {\it ordinary twitter users} and 20 {\it celebrities}. 
Experimental results demonstrate that it is feasible to 
automatically extract chronological timelines for Twitter users from their tweet collection
\footnote{Data available by request at bdlijiwei@gmail.com.}. 
\end{abstract}



\section{Introduction}
\label{intro-sec}
We have always been eager to keep track of people we are interested in. 
Entrepreneurs want to know about their competitors' past.
Employees wish to
get access to information about their boss. More generally, fans, especially young people,
are always crazy about getting the first-time news about their favorite movie stars or athletes.
To date, however, building a detailed chronological 
list or table of key events in the lives of individual remains largely 
a manual task.
Existing automatic techniques for personal event identification mostly
rely on documents produced via a web search on the person's name \cite{al2004grouping,chieu2004query,kimura2007creating,wan2005person},
and therefore are narrowed to celebrities, the lives of whom are concerned about by the online press.
While it is relatively easy to find biographical information about celebrities, as their lives are usually well documented in web, 
it is far more difficult to keep track of events in the lives of non-famous individuals.

\begin{figure}[!ht]
\centering
  \includegraphics[width=3in,natwidth=610,natheight=642]{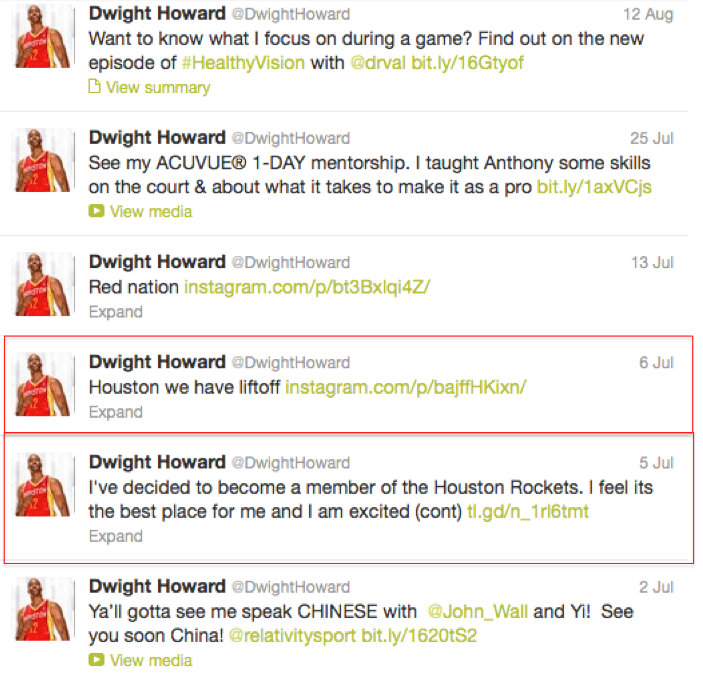}
\caption{Example of PIE for for basketball star Dwight Howard. Tweet labeled in red: PIE about joining Houston Rocket.}\label{fig1}
\end{figure}

\begin{figure}[!ht]
\centering
  \includegraphics[width=3in,natwidth=610,natheight=642]{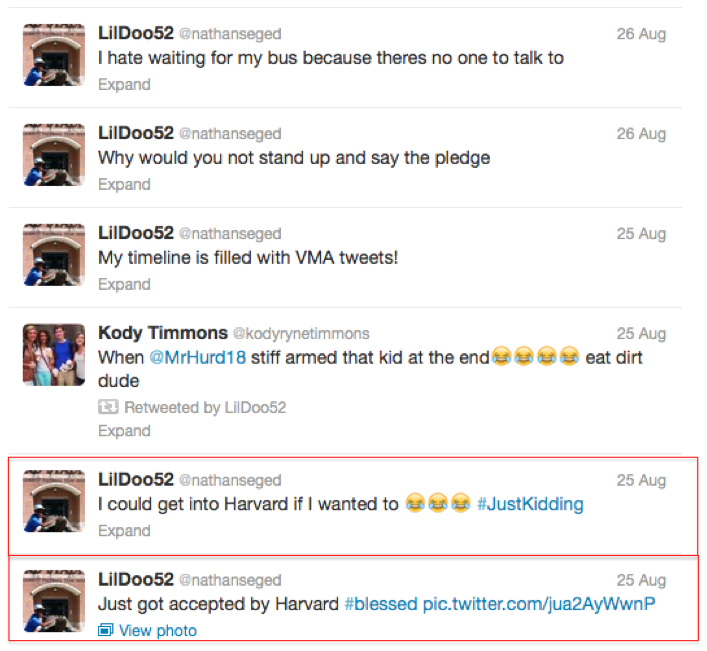}
\caption{Example of PIE for an ordinary twitter user. Tweet labeled in red: PIE about getting accepted by Harvard University. }\label{fig2}
\end{figure}

Twitter\footnote{\url{https://twitter.com/}}, a popular social
network, serves as an alternative, and potentially very rich, source of information
for this task:  people usually publish tweets describing their
lives in detail or chat with friends on Twitter as shown in Figures \ref{fig1} and \ref{fig2}, corresponding to a NBA basketball star, Dwight Howard\footnote{\url{https://twitter.com/Dwight_Howard}} tweeting about being signed by the basketball franchise Houston Rockets and an ordinary user, recording admission to Harvard University. 
Twitter provides a rich repository personal information making it amenable to automatic processing.
Can one exploit indirect clues from a relatively information-poor medium like Twitter, 
sort out important events from entirely trivial details and assemble them into an accurate life history ?

Reconstructing a person's life history from Twitter stream is an untrivial task, not only because of the extremely noisy structure that Twitter data displays, but important individual events always mixing up with entirely trivial details of little significance.
We have to answer the following question: 
what types of events reported
should be regarded as {\it personal, important events (PIE)} and therefore be
suitable for inclusion in the event timeline of an individual? 
In the current work, we specify the following three criteria for PIE extraction.  
First, a PIE should be an {\it important} event, an event that is referred to many times by an individual and his or her followers.
Second, each PIE should be a {\it time-specific} event --- a unique
(rather than a general, recurring and regularly tweeted about over a long period of time) event that is delineated by specific
start and end points.  
Consider, for instance, the twitter user in Figure \ref{fig2}, she frequently published tweets about being accepted by Harvard University
only after receiving admission notice, referring to a
time-specific PIE.  In contrast, her exercise regime, about which she tweets
regularly (e.g.\ ``11.5 km bike ride, 15 mins Yoga stretch"), is not considered a
PIE --- it is more of a general interest.


Third, the PIEs identified for an individual should be {\it personal} events
(i.e.\ an event of interest to himself or to his followers) rather than events of interest
to the general public. 
For instance, most people pay attention to and discuss about the U.S. election,
and we do not want it to be included in an ordinary person's timeline, no matter how frequently he or she tweets about it; it remains a 
public event, not a personal one.
However, things become a bit trickier because of the public nature of stardom:
sometimes an otherwise public event can constitute PIE for a celebrity --- e.g.\ ``the U.S. election" should not be treated as a PIE for ordinary individuals, but be treated as a PIE for Barack Obama and Mitt Romney.

Given the above criteria, we aim to characterize tweets into one of the following four categories: public time-specific (PublicTS), public time-general
(PublicTG), personal time-specific (PersonTS) and personal
time-general (PersonTG), as shown in Table~\ref{types-of-tweets}. 
In doing so, we can then identify PIEs related events according to the following criteria:
\begin{enumerate}
\item \textit {For an ordinary twitter user, the PIEs would the PersonTS events from correspondent user.}.
\item \textit {For a celerity, the PIEs would be PersonTS events along with his or her celebrity-related 
PublicTS events.}
\end{enumerate}

\begin{table}[tb]
\centering
 \vspace{-5pt}
\begin{tabular}{|c|c|c|}\hline
&time-specific&time-general\\\hline
public&PublicTS&PublicTG\\\hline
personal&PersonTS&PersonTG\\\hline
\end{tabular}
 \vspace{-5pt}
\caption{Types of tweets on Twitter}
\label{types-of-tweets}
 \vspace{-10pt}
\end{table}

In this paper, we demonstrate that it is feasible to 
to automatically extract chronological timelines directly from users' tweets.
In particular,
in an attempt to identify the four aforementioned types of events, 
we adapted Dirichlet Process to a multi-level representation, as we call 
Dirichlet Process Mixture Model (DPM), 
to capture
the combination of temporal information (to distinguish time-specific from time-general 
events) and user-specific information (to distinguish public from private events) in the joint Twitter feed. 
The point of DP mixture model is to allow topics shared across the data corpus while the specific level (i.e user and time) information could be emphasized. 
Further based on topic distribution from DPM model, we characterize events (topics) according to criterion mentioned above and select the tweets that best
represent PIE topics into timeline. 

For evaluation, we manually generate gold-standard PIE timelines.
As criteria for ordinary twitter users and celerities are different considering whether related PublicTS to be considered, 
we generate timelines for ordinary Twitter users denoted as $TwitSet-O$ and celebrities as $TwitSet-C$ for celebrity Twitter users respectively,
both of which include 20 users. 
In summary, this research makes the following main contributions:
\begin{itemize}
\item We create golden-standard timelines for both famous Twitter users and ordinary Twitter users
based on Twitter stream. 
\item We adapted non-parametric topic models tailored for history reconstruction for Twitter users. 
\end{itemize}

The remainder of this paper is organized as follows: 
We describe related work in Section 2 and briefly go over Dirichlet Processes 
and Hierarchical Dirichlet Processes in Section 3. 
Topic modeling technique is described in detail in Section 4 and tweet selection strategy in Section 5.
The creation of 
our dataset and gold-standards are illustrated in Section~\ref{data-sec}. 
We show experimental results in Section 7. 
We conclude with a discussion of the implications of this task to generalize and proposals for future work in Section 8.

\section{Related Work}
\label{rel-work-sec}
\paragraph{Personal Event Extraction and Timeline Generation}
Individual tracking problem can be traced back to 1996, when Plaisant et al. \cite{plaisant1996lifelines}
provided a general visualization for personal histories that can be applied to medical and court records.
Previous personal event detection mainly focus on clustering and
sorting information of a specific person from web search~\cite{al2004grouping,chieu2004query,kimura2007creating,wan2005person}.  
Existing approaches suffer from the inability of extending to average people.
The increasing popularity of
social media (e.g. twitter, Facebook\footnote{\url{https://www.facebook.com/}}), where users regularly talk about their lives, 
chat with friends, gives rise to a novel source for tracking individuals. A very similar approach is the famous Facebook Timeline\footnote{\url{https://www.facebook.com/about/timeline}}, which integrates users' status, stories and images for a concise timeline construction\footnote{The algorithm for Facebook Timeline generation remains private.}.

\paragraph{Topic Extraction from Twitter}
Topic extraction (both local and global) on twitter is not new. Among existing approaches, Bayesian topic models, either parametric (LDA \cite{blei2003latent}, labeled-LDA \cite{ramage2009labeled}) or non-parametric (HDP \cite{teh2006hierarchical}) have widely 
been widely applied due to the ability of mining latent topics hidden in tweet dataset \cite{diao2012finding,hong2010empirical,kireyev2009applications,lievolutionary,momeni2013properties,paul2011you,ramage2010characterizing,ritter2012open,zhao2011comparing}. Topic models provide a principled way to discover the topics
hidden in a text collection and seems well suited for our individual  
analysis task. 
One related work is the approach developed by Diao et al \cite{diao2012finding}, that tried to separate personal topics from public bursty topics.
Our model is inspired by earlier work that uses LDA-based topic models to separate background (generall) information 
from document-specific 
information \cite{chemudugunta2007modeling} and 
Rosen-zvi et al's work \cite{rosen2004author} that to extract user-specific information
to capture the interest of different users.



\section{DP and HDP}
\label{dp-hdp-sec}
In this section, we briefly introduce DP and HDP.
Dirichlet Process(DP) can be considered as a distribution over distributions \cite{ferguson1973bayesian}. 
A DP denoted by $DP(\alpha, G_0)$ is parameterized by a base measure $G_0$ and a concentration parameter $\alpha$.
We write $G\sim DP(\alpha,G_0)$ for a draw of distribution $G$ from the Dirichlet process. 
Sethuraman \cite{sethuraman1991constructive} showed that a measure $G$ drawn from a DP is discrete by the 
following \textit {stick-breaking construction}. 
\begin{equation}
\{\phi_k\}_{k=1}^{\infty}\sim G_0,~~~\pi\sim GEM(\alpha),~~G=\sum_{k=1}^{\infty}\pi_k\delta_{\phi_k}
\end{equation}
The discrete set of atoms $\{\phi_k\}_{k=1}^{\infty}$ are drawn from the base measure $G_0$, where $\delta_{\phi_k}$ is the probability measure concentrated at $\phi_k$.
 GEM($\alpha$) refers to the following process: 
\begin{equation}
\hat{\pi}_k\sim Beta(1,\alpha),~~~\pi_k=\hat{\pi}_k\prod_{i=1}^{k-1}(1-\hat{\pi}_i)
\end{equation}
We successively draw $\theta_1,\theta_2,...$ from measure $G$. Let $m_k$ denotes the number of draws that takes the value $\phi_k$. After observing draws $\theta_1,...,\theta_{n-1}$ from G, the posterior of G is still a DP shown as follows:
\begin{equation}
G|\theta_1,...,\theta_{n-1}\sim DP(\alpha_0+n-1,\frac{m_k\delta_{\phi_k+\alpha_0G_0}}{\alpha_0+n-1})
\label{eq3}
\end{equation}
HDP uses multiple DP s to model multiple correlated corpora. In HDP, a global measure is drawn from base measure $H$. Each document $d$ is associated with a document-specific measure 
$G_d$ which is drawn from global measure $G_0$. Such process can be summarized as follows:
\begin{equation}
G_0\sim DP(\alpha,H)~~~G_d|G_0,\gamma\sim DP(\gamma,G_0)\label{equ4}
\end{equation}
Given $G_j$, words $w$ within document $d$ are drawn from the following mixture model:
\begin{equation}
\{\theta_{w}\}\sim G_d,~~~~ w\sim Multi(w|\theta_w)\label{equ5}
\end{equation}
Eq.(\ref{equ4}) and Eq.(\ref{equ5}) together define the HDP. According to Eq.(1), $G_0$ has the form $G_0=\sum_{k=1}^{\infty}\beta_k\delta_{\phi_k}$, where $\phi_k\sim H$, $\beta\sim GEM(\alpha)$. Then $G_d$ can be constructed as
\begin{equation}
G_d\sim\sum_{k=1}^{\infty}\pi_{dk}\delta_{\phi_k},~~~\phi_j|\beta,\gamma\sim DP(\gamma,\beta)
\end{equation} 

\begin{figure}[!ht]
\centering
  \vspace{-15pt}
  \includegraphics[width=3.2in,natwidth=610,natheight=642]{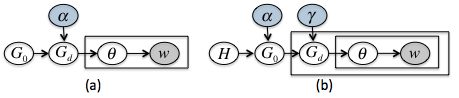}
    \vspace{-5pt}
  \caption{Graphical model of (a) DP and (b) HDP}\label{fig:dp and hdp}
  \vspace{-10pt}
\end{figure}

\section{DPM model}
\label{dpm-tracking-sec}
In this section, we get down to the details of DPM model. 
Suppose that
we collect tweets from $I$ users. Each user's tweet stream is segmented into $T$ time periods where each time period
denotes a week. $S_{i}^t=\{v_{ij}^t\}_{j=1}^{j=n_{i}^t}$ denotes the
collection of tweets that user $i$ published, retweeted or iswas
@ed during epoch $t$. Each tweet $v$ is comprised of a series of tokens $v=\{w_i\}_{i=1}^{n_v}$
where $n_v$
denotes the number of words in current tweet.
$S_i=\sum_tS_i^t$ denotes the tweet collection published by user $i$ and $S_t=\sum_iS_i^t$ denotes the tweet
collection published at time epoch t. 
$V$ is the vocabulary
size.

\subsection{DPM Model}
\label{dpm-sec}
In DPM, 
each tweet $v$ is associated with parameter $x_v$ $y_v$, $z_v$, respectively denoting whether it is Public or Personal, whether it is time-general or time-specific and the corresponding topic. We use 4 types of measures, 
each of which represents to one of PublicTG, PublicTS, PersonTG or PersonTS topics, according to
different combinations of $x$ and $y$ value. 
  
Suppose that $v$ is published by user $i$ at time $t$. $x_v$ and $y_v$ conform to the binomial distribution characterized by parameter $\pi_x^i$ and $\pi_y^i$, intended to encode users' preference for publishing personal or pubic topic, time-general or time-specific topic.  We give a Beta prior for $\pi_x^i$ and $\pi_y^i$, characterized as $\pi_x^i\sim Beta(\eta_x)$,
$\pi_y^t\sim Beta(\eta_y)$.
\begin{table}[!ht]
\centering
\begin{tabular}{|c|c|c|}\hline
&y=0&y=1\\\hline
x=0&PublicTG&PublicTS\\\hline
x=1&PersonTG&PersonTS\\\hline
\end{tabular}
\caption{Tweet type according to x (public or personal) and y (time-general or time-specific)}\label{table:type}
\end{table}

\begin{figure}[ht!]
\centering
  \vspace{-10pt}
  \includegraphics[width=3in,natwidth=610,natheight=642]{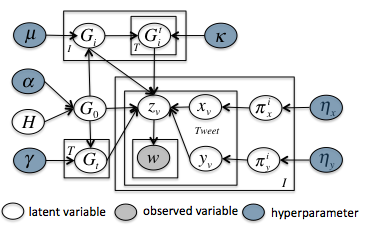}
    \vspace{-5pt}
  \caption{Graphical illustration of DPM model. }\label{fig5}
\end{figure}

In DPM, there is a global (or PublicTG) measure $G_0$, denoting topics generally talked about. A PublicTG topic (x=0, y=0) is directly drawn from $G_0$ (also denoted as $G_{(0,0)}$). $G_0$ is drawn from the base measure $H$.
For each time $t$, there is a time-specific measure $G_t$ (also denoted as $G_{(0,1)}$), which is used to characterize topics discussed specifically at time $t$ (publicTS topics).
 $G_t$ is drawn from the global measure $G_0$. Similarly, 
 for each user $i$, a user-specific $G_i$ measure (also written as $G_{(1,0)}$) is 
drawn from $G_0$ to characterize personTG topics that are specific to user $i$. 
Finally, PersonTS topic $G_i^t$ ($G_{(1,1)}$) is drawn from personal topic $G_i$ by putting a time-specific regulation. 
As we can see, all tweets from all users across all time epics share the same infinite set of mixing components (or topics). The difference lies in the mixing weights in the four types of measure $G_0$, $G_t$, $G_i$ and $G_i^t$. The whole point of DP mixture model is to allow sharing components across corpus while the specific levels (i.e., user and time) of information can be emphasized. The plate diagram and generative story are illustrated in Figures \ref{fig5} and \ref{figa}. $\alpha,\gamma,\mu$ and $\kappa$ denote hyper-parameters for Dirichlet Processes. 


\begin{figure}
\rule{8cm}{0.03cm}
\begin{tightitemize}
\item Draw PublicTG measure $G_0\sim DP(\alpha, H)$.
\item For each time t
\begin{tightitemize}
 \item draw PublicTS measure $G_t\sim DP(\gamma, G_0)$.
 \end{tightitemize}
\item For each user i
\begin{tightitemize}
\item draw $\pi_x^i\sim Beta(\eta_x)$
\item draw $\pi_y^i\sim Beta(\eta_y)$
\item draw PersonTG measure $G_i\sim DP(\mu, G_0)$.
\item for each time t:
\begin{tightitemize}
\item draw PersonTS measure $G_i^t\sim DP(\kappa, G_i)$
\end{tightitemize}
\end{tightitemize}
\item for each tweet $v$, from user i, at time t
\begin{tightitemize}
\item draw $x_v\sim Multi(\pi_x^i)$,  $y_v\sim Multi(\pi_y^i)$
\item if x=0, y=0: draw $z_v\sim G_0$
\item if x=1, y=0: draw $z_v\sim G_t$
\item if x=0, y=1: draw $z_v\sim G_i$
\item if x=1, y=1: draw $z_v\sim G_i^t$
\item for each word $w\in~v$
\begin{tightitemize}
\item draw $w\sim Multi(z_v)$
\end{tightitemize}
\end{tightitemize}
\end{tightitemize}
\rule{8cm}{0.03cm}
  \caption{Generative Story of DPM model}\label{figa}
\end{figure}

\subsection{Stick-breaking Construction}
According to the stick-breaking construction of DP, the explicit form of $G_0$, $G_i$,$G_t$,$G_i^t$ are given by:
\begin{equation}
G_0=\sum_{k=1}^{\infty}r_k\delta_{\phi_k},~~~r\sim GEM(\alpha)
\end{equation}
\begin{equation}
\begin{aligned}
&G_t=\sum_{k=1}^{\infty}\psi_t^k\delta_{\phi_k}, ~~~~\psi_t\sim DP(\gamma,r)\\
&G_i=\sum_{k=1}^{\infty}\beta_i^k\delta_{\phi_k},~~~~ \beta_i\sim DP(\mu,r)
\end{aligned}
\end{equation}
\begin{equation}
G_i^t=\sum_{k=1}^{\infty}\pi_{it}^{k}\delta_{\phi_k},~~~\pi_{it}\sim DP(\kappa,\beta_i)
\end{equation}
In this way, we obtain the stick-breaking construction for DPM, which provides a prior where $G_0$, $G_i$,$G_t$ and $G_i^t$ of all corpora at all times from all users share the same infinite topic mixture $\{\phi_k\}_{k=1}^{\infty}$. 
\subsection{Inference}
\label{inference-subsec}
In this subsection, we use Gibbs Sampling for inference. We exclude mathematical derivation for brevity, the details of which can be found in \cite{teh2006hierarchical} and \cite{zhang2010evolutionary}.

We first briefly go over Chinese restaurant metaphor for multi-level DP. A document is compared to a restaurant and the topic is compared to a dish. Each restaurant is comprised of a series of tables and each table is associated with a dish. 
The interpretation for measure $G$ in the metaphor is the dish menu denoting the list of dishes served at specific restaurant.
Each tweet is compared to a customer 
and when he comes into a restaurant, he would choose a table and shares the dish served at that table. 

{\bf Sampling r}:
What are drawn from global measure $G=\sum_{k=1}^{\infty}r_k\delta_{\phi_k}$ are the dishes for customers (tweets) labeled with (x=0, y=0) for any user across all time epoches. We denote the number of tables with dish $k$ as $M_k$ and the total number of tables as  $M_{\bullet}=\sum_{k}M_k$. Assume we already know $\{M_k\}$. Then according to Eq. (\ref{eq3}), the posterior of $G_0$ is given by:
\begin{equation}
G_0|\alpha,H,\{M_k\}\sim DP(\alpha+M_{\bullet},\frac{H+\sum_{k=1}^{\infty}M_k\delta_{\phi_k}}{\alpha+M_{\bullet}})
\end{equation}
K is the number of distinct dishes appeared. Let $Dir()$ denote Dirichlet distribution. $G_0$ can be further represented as
\begin{equation}
G_0=\sum_{k=1}^{K}r_k\delta_{\phi_k}+r_uG_u,~~~~~~r_u\sim DP(\alpha,H)\\
\end{equation}
\begin{equation}
r=(r_1,r_2,...,r_K,r_u)\sim Dir(M_1,M_2,...,M_K,\alpha)\label{r}
\end{equation}

This $augmented$ representation reformulates original infinite vector $r$ to an equivalent vector with finite-length of $K+1$
vector. $r$ is sampled from the Dirichlet distribution shown in Eq.(\ref{r}).

{\bf Sampling $\psi_t$, $\beta_i$, $\pi_{it}$}: Fraction parameters can be sampled in the similar way as $r$. 
Notably due to the specific regulatory framework for each user and time, the posterior distribution for $G_t$, $G_i$ and $G_i^t$ are calculated by only counting the number of tables in correspondent user, or time, or user and time. Take $\beta_i$ for example: as $\beta_i\sim DP(\gamma, r)$, and assume we have count variable $\{T_i^k\}$, where $T_i^k$ denotes the number of tables with topic $k$ in user $i$'s tweet corpus, the posterior for $\beta_t$ is given by:  
\begin{equation}
(\beta_i^1,...,\beta_i^K,\beta_i^u) \sim Dir(T_i^1+\gamma\cdot r_1,..., T_i^K+\gamma\cdot r_K, \gamma\cdot r_u)
\end{equation}

{\bf Sample $z_v$}: Given the value of $x_v$ and $y_v$, we sample topic assignment $z_v$ according to the correspondent $G_{x_v,y_v}$ given by:
\begin{equation}
\begin{aligned}
Pr(z_v=k|x_v,y_v,w)\propto Pr(v|x_v,y_v,z_v=k,w)\cdot Pr(z=k|G_{x,y})\\
\end{aligned}
\end{equation}
The first part $Pr(v|x_v,y_v,z_v,w)$ denotes the probability of current tweet $v$ generated by topic $z$, described in Appendix and the second part denotes the probability of dish $z$ selected from $G_{x_v,y_v}$:
\begin{equation}
\begin{aligned}
Pr(z_v=k|G_{x_v,y_v})
=\left\{
\begin{aligned}
&r_k~~(if~x=0,y=0)\\
&\psi_t^k~~(if~x=0,y=1)\\
&\beta_i^k~~(if~x=1,y=0)\\
&\pi_{i,t}^k~~(if~x=1,y=1)\\
\end{aligned}
\right.
\end{aligned}
\end{equation}

{\bf Sampling $M_k$, $T_i^k$}: Table number $M_k$, $T_i^k$ at different levels (global, user or time) of restaurants are sampled from Chinese Restaurant Process (CRP) in Teh et al.'s work \cite{teh2006hierarchical}.

{\bf Sampling $x_v$ and $y_v$}: For each tweet $v$, we determine whether it is public or personal ($x_v$), time-general or time-specific ($y_v$) as follows:  
\begin{equation}
\begin{aligned}
&Pr(x_v=x|X_{-v},v,y)\propto\frac{E_{i}^{(x,y)}+\eta_x}{E_{i}^{(\cdot,y)}+2\eta_x}\\
&~~~~~~\times \sum_{z\in G_{x,y}}P(v|x_v,y_v,z_v=k)\cdot P(z=k|G_{x,y})
\end{aligned}\label{eq10}
\end{equation}

\begin{equation}
\begin{aligned}
&Pr(y_v=y|Y_{-v},v,x)\propto\frac{E_{i}^{(x,y)}+\eta_y}{E_{i}^{(x,\cdot)}+2\eta_y}\times\\
&~~~~~~\times \sum_{z\in G_{x,y}}P(v|x_v,y_v,z_v=k,w)\cdot P(z=k|G_{x,y})
\end{aligned}\label{eq11}
\end{equation}
where $E_i^{(x,y)}$ denotes the number of tweets published by user $i$ with label (x,y) while $M_i^{(\cdot,y)}$ denotes 
number of tweets labeled as y by summing over $x$. The first part for Eqns (\ref{eq10}) and (\ref{eq11}) can be interpreted as the user's preference for publishing one of the four types of tweets while the second part as the probability of current tweet generated by the correspondent type by integrating out its containing topic $z$. 

In our experiments, we set hyperparameters $\eta_x=\eta_y=20$. The sampling for hyperparameters $\alpha,\gamma,\mu$ and $\kappa$ are decided as in Teh et al's work \cite{teh2006hierarchical} by putting a vague gamma function prior.  We run 200 burn-in iterations through all tweets to
stabilize the distribution of different parameters before collecting
samples. 
\section{Timeline Generation}
\label{timeline}
In this section, we describe how the individual timeline is generated based on DPM model. 
\subsection{Topic Merging}
Topics mined from topic models can be highly correlated \cite{lafferty2005correlated} which will lead to timeline redundancy in our task. To address this issue, we employ the hierarchical {\it agglomerative clustering} algorithm, merging mutually closest
topics into a new one step by step until the stopping conditions are met. 

The key point here is the determination of stopping conditions for the agglomerating procedure. We take strategy introduced by Jung et al.\cite{jung2003decision} that seeks the global minimum of {\it clustering balance $\varepsilon$} given by:
\begin{equation}
\varepsilon(\chi)=\Lambda(\chi)+\Omega(\chi)
\label{123}
\end{equation}
where $\Lambda$ and $\Omega$ respectively denote intra-cluster and inter-cluster error sums for a specific clustering configuration $\chi$. We use $P_{i}$ to denote the subset of topics for user $i$. 
As we can not represent each part in Equ.\ref{123} as Euclidean Distance as in the case of standard clustering algorithms,
we adopt the following probability based distance metrics: we use entropy to represent intra-cluster error given by:
 \begin{equation}
\Lambda(P_i)=\sum_{L\in P_i}\sum_{v\in L}-p(v|C_l)\log p(v|C_l)
\end{equation}
\noindent 
The inter-cluster error $\Omega$ is measured by the KL divergence between each topic and the topic center $C_{P_i}$:
\begin{equation}
\Omega(C_i)=\sum_{L\in P_i}KL(L||C_{P_i})
\end{equation}

Stopping condition is achieved when the minimum value of {\it clustering balance} is obtained. 

\subsection{Selecting Celerity related PublicTS}
In an attempt to identify celebrity related PublicTS topics, we employ rules
based on (1) user name co-appearance (2) p-value for topic {\it shape  comparison} and (3) {\it clustering balance}.
For a celebrity user $i$, 
a PublicTS topic $L_j$ would be considered 
as a celebrity related if it satisfies:
\begin{enumerate}
\item user $i$'s name or twitter $id$ appears in at least $10\%$ of tweets belonging to $L_j$.
\item The $P-value$ for $\chi^2$ {\it shape comparison} between $G_i$ and $L_j$ is larger than 0.5. 
\item $\varepsilon\{D\bigcup L_j, L_1, L_2, ..., L_{j-1}, L_{j+1},...\}\leq \varepsilon\{D,L_1,L_2,...L_j, ...\}$. 
\end{enumerate}

\subsection{Tweet Selection}
The tweet that best represents the PIE topic $L$ is selected into timeline:
\begin{equation}
v_{select}=\argmax_{v'\in L}Pr(v'|L)
\end{equation}

\section{Data Set Creation}
\label{data-sec}
We describe the creation of our Twitter data set and gold-standard PIE timelines used to train
and evaluate our models in this Section. 

\subsection{Twitter Data Set Creation}
\label{twitter-data-subsec}
Construction of the DPM model (as well as the baselines) requires 
the tweets of both famous and non-famous people.
We crawled about 400 million tweets from 500,000 users from Jun 7th, 2011 through Mar 4th, 2013, 
from Twitter API\footnote{\url{http://dev.twitter.com/docs/streaming-api}}.
The time span
totals 637 days, which we split into 91 time periods (weeks).

From this set, 
we identify 20 ordinary users with the number of followers between 500 and 2000 and publishing more than 1000 tweets within the designated time period and crawled 36, 520 from their Twitter user accounts.
We further identify 20 celebrities (details see Section 6.2) as Twitter users with more than 1,000,000 followers.
Due to Twitter API limit\footnote{one can crawl at most 3200 tweets for each user}, we also harness data from CMU Gardenhose/Decahose which contains roughly $10\%$ of all Twitter postings.
We fetch tweets containing @ specific user\footnote{For celebrities, the number tweets with @ are much greater than their published tweets.} from Gardenhose. 
The resulting data set contains 132,423 tweets for the 20 celebrities.

For simplicity, instead of pulling all tweet-containing tokens into DPM model, we represent each tweet with its containing nouns and verbs. Part of Speech tags are assigned based on Owoputi et al's tweet POS system \cite{owoputi2013improved}.
Stop-words are removed.

\subsection{Gold-Standard Dataset Creation}
\label{timeline-subsec}

For evaluation purposes, we respectively generate gold-standard PIE dataset for ordinary twitter users and celebrities separately based on 
one's Twitter stream.
\paragraph{Twitter Timeline for Ordinary Users ($TwitSet-O$)} 
To generate golden-standard timeline for ordinary users, we chose 20 different Twitter users. 
In a sense that no one understands your true self better than you do, we asked them to identify each of his or her tweets as either PIE related according to their own experience. 
In addition, each PIE-tweet is labeled with a short
string designating a name for the associated PIE. Note that multiple tweets can be labeled with the same event name. For ordinary user gold-standard generation, we only ask the user himself for labeling and no inter-annotator agreement is measured. This is reasonable considering the reliability of user labeling his own tweets.

\begin{table}
\small
\centering
\begin{tabular}{cccc}\hline
Name&TwitterID&Name&TwitterID\\\hline
Lebron James&KingJames&Ashton Kutcher&aplusk\\
Dwight Howard&DwightHoward&Russell Crowe&russellcrowe\\
Serena Williams&serenawilliams&Barack Obama&BarackObama\\
Rupert Grint&rupertgrintnet&Novak Djokovi&DjokerNole\\
Katy Perry&katyperry&Taylor Swift&taylorswift13\\
Dwight Howard&DwightHoward&Jennifer Lopez&JLo\\
Wiz Khalifa&wizkhalifa&Chris Brown&chrisbrown\\
Mariah Carey&MariahCarey&Kobe Bryant&kobebryant\\
Harry Styles&Harry-Styles&Bruno Mars&BrunoMars\\
Alicia Keys&aliciakeys&Ryan Seacrest&RyanSeacrest\\\hline
\end{tabular}
\caption{List of Celebrities in $TwitSet-C$.}\label{table:names}
\end{table}

\begin{figure}[!ht]
\centering
\includegraphics[width=3.4in]{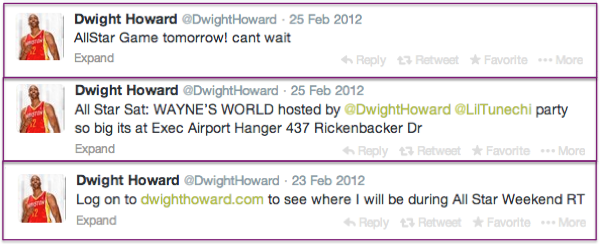}
\includegraphics[width=3.4in]{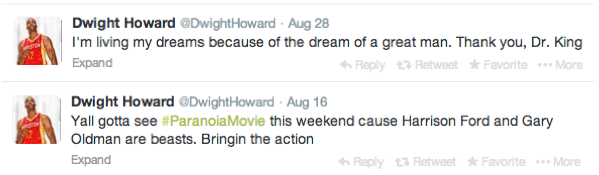}
\includegraphics[width=3.4in]{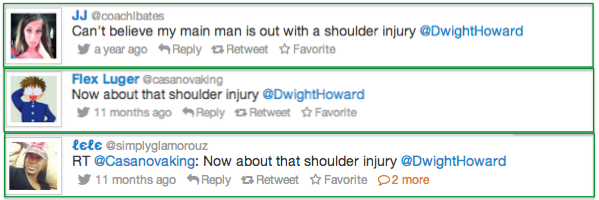}
\includegraphics[width=3.4in]{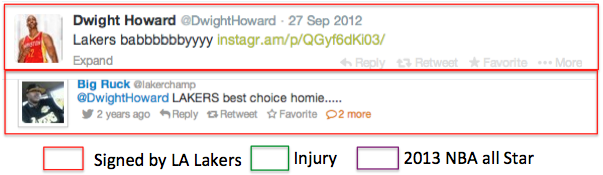}
\caption{Example of Golden-Standard timeline $TwitSet-C$ for basketball star Dwight Howard.  Colorful rectangles denote PIEs given by annotators. Labels are manually given.}\label{twitter set}
\end{figure}

\paragraph{Twitter Timeline for CelebritY Users ($TwitSet-C$)} 
We first employed workers from Amazon's Mechanical Turk\footnote{\url{https://www.mturk.com/mturkE/welcome}} to label each tweet in $TwitSet-O$ as PIE-related or not PIE-related (shown in Table \ref{table:names}). 
We assigned each tweet to 2 different workers. $Cohen~~\kappa$ is used to measure inter-agreement. 
Unfortunately, 
the average value for $\kappa$ is 0.653 with standard deviation 0.075 in the evaluation, not showing substantial agreement. To address this issue, we further turned to the crowdsourcing service {\it oDesk}\footnote{\url{https://www.odesk.com/}}, which allows requesters to recruit individual workers with specific skills. We recruited two workers for each celebrity based on their ability to answer certain questions on related fields, say ``who is the MVP for NBA regular season 2011" when labeling NBA basketball stars (i.e. Dwight Howard, Lebron James) or ``at which year Russell Crowe's movie {\it Gladiator} won him Oscar best actor" when labeling Russell Crowe. 
More specialized and experienced workers would generate better gold-standards. 
These experts in $oDesk$ agree with a $\kappa$ score of 0.901, showing substantial agreement. For the small amount of labels on which the judges disagree, we recruited an extra judge and to serve as a tie breaker. Illustration for the generation of $TwitSet-C$ is shown in Figure \ref{twitter set}. 

\begin{table}
\small
\centering
\begin{tabular}{ |c|c|c|c| }
\hline
&TwitSet-O&TwitSet-C\\\hline
$\#$ of PIEs&112&221\\\hline
avg $\#$ PIEs per person&6.1& 11.7\\\hline
max $\#$ PIEs per person&14& 20\\\hline
min $\#$ PIEs per person&2&3\\\hline
\end{tabular}
\caption{Statistics for Gold-standard timelines}\label{table:set}
\end{table}


\section{Experiments}
\label{experiments}
In this section, we evaluate our approach to PIE timeline construction for 
both ordinary users and famous users by comparing the results of DPM with baselines. 

\subsection{Baselines}
We implement the following baselines for comparison.
We use identical 
processing techniques for each approach for fairness. 

\begin{figure}[!ht]
\includegraphics[width=3.2in]{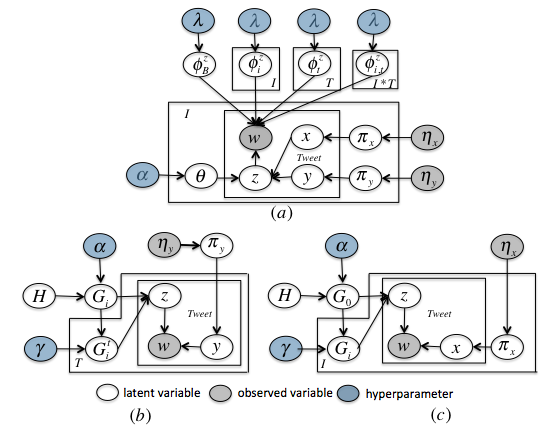}
\caption{Graphical illustration for baselines: (a) Multi-level LDA (b) Person-DP (c) Public-DP. \label{10}}
\end{figure}

{\bf Multi-level LDA}: Multi-level LDA is similar to HDM but uses LDA based topic approach (shown in Figure \ref{10}(a)) for topic mining. Latent parameter $x_v$ and $y_v$ are used to denote 
whether the correspondent tweet is personal or public, time-general or time-specific. 
Different combinations of $x$ and $y$ is characterized by different distributions over vocabularies: a background topic $\phi_B$ for (x=0,y=0), time-specific topic for $\phi_t$ for (x=0, y=1), user-specific topic $\phi_i$ for (x=1,y=0) and time-user-specific topic $\phi_{i}^t$ for (x=1,y=1).

{\bf Person-DP}: A simple version of DPM model that takes only as input tweet stream published by one specific user, as shown in Figure \ref{10}(b). Consider one particular Twitter user $i$,
{\it Person-DP} aims at separating his time-specific topic $G_i^t$,
from background topic $G_i$.

{\bf Public-DP:} A simple version of DPM that separates personal topics $G_i$
from public events/topics $G_0$ as shown in Figure \ref{10}(c).

\begin{table}
\centering
\begin{tabular}{|c|c|c|c|}\hline
PublicTG&PublicTS&PersonTG&PersonTS\\\hline
39.7$\%$&20.3$\%$&21.2$\%$&18.8$\%$\\\hline
\end{tabular}
\caption{Percentage of different types of Tweets.}\label{table6}
\end{table}

\subsection{Results for PIE Timeline Construction}
Performance on the central task of identifying the personal important events
of celebrities is the {\bf Event-level Recall}, as shown in 
Table~6, which shows the percentage of PIEs from
the Twitter-based gold-standard timeline that each model can retrieve.
One PIE is regarded as retrieved if at least one of the event-related tweets is correctly identified.

As we can see from Table~6, the recall rate for $Twit-C$ is much higher than $Twit-O$. 
As celebrities tend to have more followers, 
their PIE related tweets are usually followed, retweeted and replied by great number of followers.
Even for those PIEs that can not be evidently discovered from user's own Twitter collection, postings from followers
can provide strong evidence.  
For baseline comparison, 
$DPM$ is a little bit better than $Multi-level~LDA$ due to its non-parametric nature and ability in modeling topics shared across the 
corpus. Notably, {\it Public-DP} obtains the best recall rate for {\it Twit-O}. The reason is that {\it Public-DP} includes all personal information into the timeline, regardless of whether it is time-general or time-specific. The high recall rate of {\it Public-DP}
is sacrificed extremely by low precision rate
as we will talk about below. 

\begin{table}[!ht]
\centering
\small
\begin{tabular}{|c|c|c|}\hline
\multirow{3}{*}{Approach}&\multicolumn{2}{c|}{Event-level Recall}\\\hline
&$Twit-O$&$Twit-C$\\\hline
DPM&0.752&{\bf 0.927}\\\hline
Multi-level LDA&0.736&0.882\\\hline
Person-DP&0.683&0.786\\\hline
Public-DP&{\bf 0.764}&0.889\\\hline
\end{tabular}
\caption{Evaluation for different systems.}\label{table:systems evaluate}
\end{table}

\begin{table*}[!ht]
\centering
\small
\begin{tabular}{|c|c|c|c|c|c|c| }\hline
\multirow{2}{*}{Approach}&\multicolumn{3}{c}{$Twit-O$}&
\multicolumn{3}{|c|}{$Twit-C$}\\\cline{2-7}
&Precision&Recall&F1&Precision&Recall&F1\\\hline
DPM&{\bf 0.798}&0.700&{\bf 0.742}&{\bf 0.841}&0.820&0.830\\\hline
Multi-level LDA&0.770&0.685&0.725&0.835&0.819&0.827\\\hline
Person-DP&0.562&0.636&0.597&0.510&0.740&0.604\\\hline
Public-DP&0.536&{\bf 0.730}&0.618&0.547&{\bf 0.823}&0.657\\\hline
\end{tabular}
\caption{Tweet level result evaluation for different systems.}\label{table:systems evaluate}
\vspace{-5pt}
\end{table*}

\subsection{Results for Tweet-Level Prediction}
Although our main concern is the percentage of PIEs each model can retrieve, the precision for tweet-level predictions of each model is
also potentially of interest. There is a \textsc{trade-off} between the event-level recall and the tweet-level precision as more tweets means more topics being covered, but more likely non-PIE-related tweets included as well.
We report Precision, Recall and F-1 scores regarding whether a PIE tweet is identified in Table \ref{table:systems evaluate}.

As we can observe from Table~\ref{table:systems evaluate}, $DPM$ and $Multi-level~LDA$  outperform other baselines by a large margin with respect to Tweet-Level precision rate.
As {\it Personal-DP} takes as input tweet collection from single user and does not distinguish between personal and public topics, events such as American Presidential Election concerned by individual users will be mis-classified, leading to low precision score. 
{\it Public-DP} does not distinguish between PersonTG and PersonTS topics and therefore includes reoccurring topics into timeline, and therefore gets low precision score as well.

\begin{table*}[!ht]
\small
\centering
\begin{tabular}{|l|p{10cm}|p{3cm}|}\hline
Time Periods&Tweets selected by DPM&Manually Label\\\hline
Jun.13 2011&Enough of the @KingJames hating. He's a great player. And great players don't play to lose, or enjoy losing, even it is final game.&2011 NBA Finals\\\hline
Jan.01 2012&AWww @kingjames is finally engaged!! Now he can say hey kobe I have ring too :P.&Engaged \\\hline
Feb.20 2012&We rolling in deep to All-Star weekend, let's go&2012 NBA All-Star.\\\hline
Jun.19 2012&OMFG I think it just hit me, I'm a champion!! I am a champion!&2012 NBA Finals \\\hline
Jul.01 2012&$\#$HeatNation please welcome our newest teammate Ray Allen.Wow&
Welcome Ray Allen\\\hline
Jul.15 2012&@KingJames won the award for best male athlete.&Best Athlete award\\\hline
Aug.05 2012&What a great pic from tonight @usabasketball win in Olypmic!&Win Olympics\\\hline
Sep.02 2012& {\slshape Big time win for @dallascowboys in a tough environment tonight! Great start to season. Game ball goes to Kevin Ogletree}&Wrongly detected\\\hline
\end{tabular}
\caption{Chronological table for Lebron James from DPM.}\label{table:james}
\end{table*}

\subsection{Sample Results and Discussion}
In this subsection, we present part of sample results outputted. Table \ref{table6} presents percentage of different types of tweets according to DPM model. PublicTG takes up the largest portion, up to about $38\%$. PublicTG is followed by PublicTS and PersonTG topics and then PersonTS.

Next, we present PIE related topics extracted from an 22-year-old female Twitter user who is a senior undergraduate from Cornell University and a famous NBA basketball player, Lebron James. Correspondent top words within the topics are presented in Tables 9 and 10. 
As we can see from Table 9, for the ordinary user, the 4 topics mined respectively correspond to (1) her internship at Roland Beger in Germany (2) The role in played in the drama ``A Midsummer Night's Dream" (3) Graduation (4) Starting a new job at BCG, New York City. 
For Labron James, the topics correspond to (1) NBA finals (2) Ray Allen joined basketball franchise Heat (3) 2012 Olympics and (4) NBA All-Star game.
Topic labels are manually given. 
One interesting direction for future work is automatic labeling PIE events detected and generating a much conciser timeline based on automatic labels \cite{lau2011automatic}.
 Table~\ref{table:james} shows the timeline generated by our system for Lebron James. We can clearly observe PIE related events such as NBA all-Star, NBA finals or being engaged can be well detected. The tweet in italic font is a wrongly detected PIE. This tweet talks about the DallasCowboys, a football team which James is interested in. It should be regarded as an interest rather than a PIE. 
James published a lot of tweets about DallasCowboys during a very short period of time and they are wrongly treated as PIE related. 

\begin{table}[!ht]
\small
\centering 

          \begin{tabular}{|c|c|c|}\hline
          &manual label&top word\\\hline
          Topic 1&summer intern&intern, Roland\\
          &&Berger, berlin\\\hline
          Topic 2&play a role in drama&Midsummer, Hall, act\\
          &&cheers, Hippolyta\\\hline
          Topic 3&graduation&farewell,Cornell\\
          &&ceremony,prom\\\hline
          Topic 4&begin working&York, BCG\\
          &&NYC,office\\\hline
          \end{tabular}
          \caption{Top words of topics from the ordinary twitter user}
          \end{table}
          
\begin{table}[!ht]
\small
\centering
                    \begin{tabular}{|c|c|c|}\hline
          &manual label&top word\\\hline
          Topic 1&NBA 2012 finals&finals, champion,Heat\\
          &&Lebron, OKC\\\hline
          Topic 2&Ray Allen join Heat&Allan, Miami, sign\\
          &&Heat, welcome\\\hline
          Topic 3&2012 Olympic&Basketball, Olympic\\
          &&Kobe, London, win\\\hline
          Topic 4&NBA All-Star Game&All-star, Houston\\\hline
          \end{tabular}
          \caption{Top words of topics from Lebron James}                
                    \label{table1000}
\end{table}

\section{Conclusion, Discussion and Future Work}
\label{conclusion-sec}
In this paper, we preliminarily study the problem of individual timeline generation problem for Twitter users and 
propose
a tailored algorithm for {\it personal-important-event} (PIE) identification
by distinguishing four
types of tweets, namely PublicTS, PublicTG, PersonPS and
PersonPG. 
Our algorithm is predicated on the assumption that PIE related topics should be both personal (opposite to public) and time-specific (opposite to time-specific). While our approach enjoys good performance on the tested data,
it suffers from the following disadvantages:

First, there are both gains and losses with the unsupervised nature of our model. 
The gains are that it frees us from the great difficulties in obtaining gold-standard labeled data in this task. 
However, topic models harness word frequency as features for topic modeling, which means a topic must be 
adequately talked about to ensure it to be discovered. Results in Section \ref{experiments} also demonstrate this point where performance on celebrity dataset outperforms ordinary user dataset, as topics for celebrities are usually adequately discussed.
Many average users in real life maintain a low profile on Twitter: they do not regularly update their status or do not have great number of followers to enable personal topics substantially discussed.
In that case, topic models would fail to work. For example, if no one replies one's posting about admission to some univeristy, 
it would be hard for topic model based approach to retrieve such PIE topic and include it in the timeline.

Second, the time-specific assumption is not permanent-perfect. For example, after one gets into Harvard University for undergraduate study, he tends to frequently tweet about the college he is affiliated with. In that case, the keyword ``Harvard"  changes from a time-specific word to a time-general one. The time-specific assumption may confuse the two situations and fail to list the acceptance in the timeline.
Additionally, the time-specific concept can not distinguish between short-term interests and PIE topics, as shown in Lebron James's example in Section 7.2.

Our future work constitutes combining both supervised and unsupervised algorithms that promises better timeline generation. 
One direction would be using weak (or distant) supervision (e.g., \cite{hoffmann2011knowledge,mintz2009distant}) where training data can be retrieved by matching tweets to ground truths from external sources, such as Facebook or Wikipedia. 
Notably, Facebook supports individual timeline application and seems as a good fit for this task. 
Another promising perspective is either manually or automatically constructing a comprehensive list of categories about individual PIEs in the first place, such as education, job, marriage, travel, and then use the list as guidelines for later timeline construction. Additionally, our system is inherently more of a tweet selection approach than a timeline \textsc{generation} algorithm. An individual history comprised of raw tweet data is poorly readable. Integrating summarization techniques for a better PIE representation also constitutes our future work.

It is also worth noting that automatic individual history extraction 
may raise privacy concerns. Although Twitter feeds are public by design, the idea of a person's personal history being easily retrieved or analyzed by others  
may not be the one that is welcomed by every Twitter user.

\section{Acknowledgement}
We thank Myle Ott, Sujian Li, Alan Ritter, Wenjie Li and Chris Dyer for their insightful discussions and suggestions.
Katia Sycara, Xun Wang and Tao Ge helped with the original version.
This work was supported in part by the Language Technology Institute of Carnegie Mellon University, 
National Science Foundation Grant BCS-0904822,
a DARPA Deft grant,
as well as a gift from Google.

\bibliographystyle{abbrv}
\bibliography{sigproc}

\begin{thebibliography}{10}

\bibitem{al2004grouping}
R.~Al-Kamha and D.~W. Embley.
\newblock Grouping search-engine returned citations for person-name queries.
\newblock In {\em Proceedings of the 6th annual ACM international workshop on
  Web information and data management}, pages 96--103. ACM, 2004.

\bibitem{blei2003latent}
D.~M. Blei, A.~Y. Ng, and M.~I. Jordan.
\newblock Latent dirichlet allocation.
\newblock {\em the Journal of machine Learning research}, 3:993--1022, 2003.

\bibitem{chemudugunta2007modeling}
C.~Chemudugunta and P.~S.~M. Steyvers.
\newblock Modeling general and specific aspects of documents with a
  probabilistic topic model.
\newblock In {\em Advances in Neural Information Processing Systems:
  Proceedings of the 2006 Conference}, volume~19, page 241. The MIT Press,
  2007.

\bibitem{chieu2004query}
H.~L. Chieu and Y.~K. Lee.
\newblock Query based event extraction along a timeline.
\newblock In {\em Proceedings of the 27th annual international ACM SIGIR
  conference on Research and development in information retrieval}, pages
  425--432. ACM, 2004.

\bibitem{diao2012finding}
Q.~Diao, J.~Jiang, F.~Zhu, and E.-P. Lim.
\newblock Finding bursty topics from microblogs.
\newblock In {\em Proceedings of the 50th Annual Meeting of the Association for
  Computational Linguistics: Long Papers-Volume 1}, pages 536--544. Association
  for Computational Linguistics, 2012.

\bibitem{ferguson1973bayesian}
T.~S. Ferguson.
\newblock A bayesian analysis of some nonparametric problems.
\newblock {\em The annals of statistics}, pages 209--230, 1973.

\bibitem{hoffmann2011knowledge}
R.~Hoffmann, C.~Zhang, X.~Ling, L.~S. Zettlemoyer, and D.~S. Weld.
\newblock Knowledge-based weak supervision for information extraction of
  overlapping relations.
\newblock In {\em ACL}, pages 541--550, 2011.

\bibitem{hong2010empirical}
L.~Hong and B.~D. Davison.
\newblock Empirical study of topic modeling in twitter.
\newblock In {\em Proceedings of the First Workshop on Social Media Analytics},
  pages 80--88. ACM, 2010.

\bibitem{jung2003decision}
Y.~Jung, H.~Park, D.-Z. Du, and B.~L. Drake.
\newblock A decision criterion for the optimal number of clusters in
  hierarchical clustering.
\newblock {\em Journal of Global Optimization}, 25(1):91--111, 2003.

\bibitem{kimura2007creating}
R.~Kimura, S.~Oyama, H.~Toda, and K.~Tanaka.
\newblock Creating personal histories from the web using namesake
  disambiguation and event extraction.
\newblock In {\em Web Engineering}, pages 400--414. Springer, 2007.

\bibitem{kireyev2009applications}
K.~Kireyev, L.~Palen, and K.~Anderson.
\newblock Applications of topics models to analysis of disaster-related twitter
  data.
\newblock In {\em NIPS Workshop on Applications for Topic Models: Text and
  Beyond}, volume~1, 2009.

\bibitem{lafferty2005correlated}
J.~D. Lafferty and D.~M. Blei.
\newblock Correlated topic models.
\newblock In {\em Advances in neural information processing systems}, pages
  147--154, 2005.

\bibitem{lau2011automatic}
J.~H. Lau, K.~Grieser, D.~Newman, and T.~Baldwin.
\newblock Automatic labelling of topic models.
\newblock In {\em ACL}, volume 2011, pages 1536--1545, 2011.

\bibitem{lievolutionary}
J.~Li and S.~Li.
\newblock Evolutionary hierarchical dirichlet process for timeline
  summarization.
\newblock 2013.

\bibitem{mintz2009distant}
M.~Mintz, S.~Bills, R.~Snow, and D.~Jurafsky.
\newblock Distant supervision for relation extraction without labeled data.
\newblock In {\em Proceedings of the Joint Conference of the 47th Annual
  Meeting of the ACL and the 4th International Joint Conference on Natural
  Language Processing of the AFNLP: Volume 2-Volume 2}, pages 1003--1011.
  Association for Computational Linguistics, 2009.

\bibitem{momeni2013properties}
E.~Momeni, C.~Cardie, and M.~Ott.
\newblock Properties, prediction, and prevalence of useful user-generated
  comments for descriptive annotation of social media objects.
\newblock In {\em Seventh International AAAI Conference on Weblogs and Social
  Media}, 2013.

\bibitem{owoputi2013improved}
O.~Owoputi, B.~O'Connor, C.~Dyer, K.~Gimpel, N.~Schneider, and N.~A. Smith.
\newblock Improved part-of-speech tagging for online conversational text with
  word clusters.
\newblock In {\em Proceedings of NAACL-HLT}, pages 380--390, 2013.

\bibitem{paul2011you}
M.~J. Paul and M.~Dredze.
\newblock You are what you tweet: Analyzing twitter for public health.
\newblock In {\em ICWSM}, 2011.

\bibitem{plaisant1996lifelines}
C.~Plaisant, B.~Milash, A.~Rose, S.~Widoff, and B.~Shneiderman.
\newblock Lifelines: visualizing personal histories.
\newblock In {\em Proceedings of the SIGCHI conference on Human factors in
  computing systems}, pages 221--227. ACM, 1996.

\bibitem{ramage2010characterizing}
D.~Ramage, S.~T. Dumais, and D.~J. Liebling.
\newblock Characterizing microblogs with topic models.
\newblock In {\em ICWSM}, 2010.

\bibitem{ramage2009labeled}
D.~Ramage, D.~Hall, R.~Nallapati, and C.~D. Manning.
\newblock Labeled lda: A supervised topic model for credit attribution in
  multi-labeled corpora.
\newblock In {\em Proceedings of the 2009 Conference on Empirical Methods in
  Natural Language Processing: Volume 1-Volume 1}, pages 248--256. Association
  for Computational Linguistics, 2009.

\bibitem{ritter2012open}
A.~Ritter, O.~Etzioni, S.~Clark, et~al.
\newblock Open domain event extraction from twitter.
\newblock In {\em Proceedings of the 18th ACM SIGKDD international conference
  on Knowledge discovery and data mining}, pages 1104--1112. ACM, 2012.

\bibitem{rosen2004author}
M.~Rosen-Zvi, T.~Griffiths, M.~Steyvers, and P.~Smyth.
\newblock The author-topic model for authors and documents.
\newblock In {\em Proceedings of the 20th conference on Uncertainty in
  artificial intelligence}, pages 487--494. AUAI Press, 2004.

\bibitem{sethuraman1991constructive}
J.~Sethuraman.
\newblock A constructive definition of dirichlet priors.
\newblock Technical report, DTIC Document, 1991.

\bibitem{teh2006hierarchical}
Y.~W. Teh, M.~I. Jordan, M.~J. Beal, and D.~M. Blei.
\newblock Hierarchical dirichlet processes.
\newblock {\em Journal of the american statistical association}, 101(476),
  2006.

\bibitem{wan2005person}
X.~Wan, J.~Gao, M.~Li, and B.~Ding.
\newblock Person resolution in person search results: Webhawk.
\newblock In {\em Proceedings of the 14th ACM international conference on
  Information and knowledge management}, pages 163--170. ACM, 2005.

\bibitem{zhang2010evolutionary}
J.~Zhang, Y.~Song, C.~Zhang, and S.~Liu.
\newblock Evolutionary hierarchical dirichlet processes for multiple correlated
  time-varying corpora.
\newblock In {\em Proceedings of the 16th ACM SIGKDD international conference
  on Knowledge discovery and data mining}, pages 1079--1088. ACM, 2010.

\bibitem{zhao2011comparing}
W.~X. Zhao, J.~Jiang, J.~Weng, J.~He, E.-P. Lim, H.~Yan, and X.~Li.
\newblock Comparing twitter and traditional media using topic models.
\newblock In {\em Advances in Information Retrieval}, pages 338--349. Springer,
  2011.

\end{thebibliography}

\appendix
\label{Appendix}
\section{Calculation of f(v|x,y,z)}
\label{f}
$Pr(v|x,y,x,z,w)$ denotes the probability that current tweet is generated by an
existing topic $z$ and $Pr(v|x,y,z_{new},w)$ denotes the probability current tweet is generated by the new topic.  Let
$E_{(z)}^{(\cdot)}$ denote the number of words assigned to topic $z$
in tweet type $x,y$ and $E_{(z)}^{(w)}$ denote the number of
replicates of word $w$ in topic $z$. $N_{v}$ is the number of words
in current tweet and $N_{v}^{w}$ denotes of replicates of word $w$
in current tweet. We have:
\begin{equation*}
Pr(v|x,y,z,w)=\frac{\Gamma(E_{(z)}^{(\cdot)}+V\lambda)}{\Gamma(E_{(z)}^{(\cdot)}+N_{v}+V\lambda)}
\cdot \prod_{w\in v}\frac{\Gamma(E_{(z)}^{(w)}+N_{v}^{w}+\lambda)}{\Gamma(E_{(z)}^{(w)}+\lambda)}
\end{equation*}
\begin{equation*}
Pr(v|x,y,z^{new},w)=\frac{\Gamma(V\lambda)}{\Gamma(N_{v}+V\lambda)}
\cdot \prod_{w\in v}\frac{\Gamma(N_{v}^{w}+\lambda)}{\Gamma(\lambda)}
\end{equation*}
 $\Gamma( )$ denotes gamma function and $\lambda$ is the Dirichlet prior, set to 0.1. 

\end{document}